\documentclass[conference]{IEEEtran}
\IEEEoverridecommandlockouts
\usepackage[utf8]{inputenc}
\usepackage{cite}
\usepackage{floatrow}
\usepackage{amsmath,amssymb,amsfonts,stfloats}
\usepackage{tabularx}
\usepackage{algorithmic}
\usepackage{graphicx}
\usepackage{textcomp}
\usepackage{floatrow}
\usepackage{xcolor}
\usepackage{booktabs}

\usepackage{multicol}
\usepackage{acronym}
\usepackage{url}
\usepackage{mathtools}
\usepackage[font=small]{caption}
\usepackage[caption=false,font=footnotesize]{subfig}
\usepackage{comment}
\def\BibTeX{{\rm B\kern-.05em{\sc i\kern-.025em b}\kern-.08em
    T\kern-.1667em\lower.7ex\hbox{E}\kern-.125emX}}

\newcommand{\linebreakand}{%
  \end{@IEEEauthorhalign}
  \hfill\mbox{}\par
  \mbox{}\hfill\begin{@IEEEauthorhalign}
}

\begin{document}

\title{Multi-View Near-field Imaging in NLOS with Non-Reconfigurable EM Skins
\thanks{This work was partially supported by the European Union under the Italian National Recovery and Resilience Plan (NRRP) of NextGenerationEU, partnership on “Telecommunications of the Future” (PE00000001 - program “RESTART”).}
}
\author{Davide Tornielli Bellini,  Dario~Tagliaferri, Marouan~Mizmizi, Stefano~Tebaldini and Umberto~Spagnolini\\
Politecnico di Milano, Milan, Italy\\
E-mail:\,name.surname@polimi.it}


\maketitle

\begin{abstract}
This paper deals with radar imaging in non-line of sight (NLOS) with the aid of non-reconfigurable electromagnetic skins (NR-EMSs). NR-EMSs are passive metasurfaces whose reflection properties are defined during the manufacturing process, and represent a low-cost alternative to reconfigurable intelligent surfaces to implement advanced wave manipulations. We propose and discuss a multi-view near-field radar imaging system where a moving source progressively illuminates different portions of the NR-EMS, whereby each portion (\textit{module}) is purposely phase-configured to focus the impinging radiation over a desired NLOS area of interest. The source, e.g., a radar-equipped vehicle, synthesizes a wide aperture that maps onto the NR-EMS, allowing NLOS imaging with enhanced resolution compared to the standalone radar capabilities. Simulation results show the feasibility and benefits of such an imaging approach and shed light on a possible practical application of metasurfaces for sensing. 
\end{abstract}

\begin{IEEEkeywords}
Radar imaging, multi-view, NLOS, near-field, EM skins, synthetic aperture
\end{IEEEkeywords}

\section{Introduction}
Radar sensing technology has witnessed remarkable progress in recent years, driven by the constant pursuit of improving overall performance. Among the innovative approaches, the integration of metasurfaces into radar systems has emerged as a promising frontier. Metasurfaces are planar or conformal metamaterial surfaces whose EM properties can be can be engineered to implement advanced wave manipulation (anomalous reflection, focusing, beam shaping, etc. \cite{doi:10.1126/science.1210713}). When manufactured as reflectarrays, metasurfaces are made by a very large number of sub-wavelength elements (\textit{meta-atoms}), whose scattering phases are designed to accomplish the aforementioned functionalities. If the phase across the metasurface is time-varying, we refer to reconfigurable intelligent surfaces (RISs), and their application to communication and localization systems has been the focus of the vast majority of research works. In~\cite{Buzzi_RISforradar_journal}, for instance, the authors propose to use a RIS to assist radar in non-line-of-sight (NLOS) conditions. The work~\cite{9511765}, instead, uses a RIS for NLOS radar localization of targets. The authors of \cite{9508872} analyze the fundamental position and orientation error bounds on a RIS-aided localization system, proposing a suitable phase design approach. Recently, the research interest shifted towards large metasurfaces (whose size is comparable with the propagation distance), to exploit the additional degrees of freedom brought by \textit{near-field} operation, namely the propagation condition in which the wavefront across the metasurface is non-planar~\cite{9838638}. Examples of works on RIS-aided near-field localization are in \cite{9625826,9650561}. 

Differently from localization, whose aim is to infer meaningful parameters of a set of desired targets (e.g., position, orientation, speed, etc.), \textit{imaging} refers to the generation of a reflectivity map of the environment, from which to infer the presence of targets via detection, prior to localization. Imaging performance is not assessed in terms of estimation accuracy but rather resolution, i.e., the capability of distinguishing two closely spaced targets. Currently, very few works address the usage of RISs (and metasurfaces in general) for imaging (e.g., see \cite{PhysRevLett.92.193904}), with specific use cases mostly limited to biomedical imaging \cite{Ghavami2022_metasurfaces_radar}. Other relevant works are \cite{9299878,jiang2023near,torcolacci2023holographic}. The authors of \cite{9299878} address the imaging problem as an inverse scattering reconstruction in far-field, considering distributed antenna systems. Work \cite{jiang2023near}, instead, addresses the near-field by employing the RIS as an active reflector, amplifying and reflecting the impinging signals from a large holographic aperture. Recently, the authors of \cite{torcolacci2023holographic} tackled the imaging with RISs in both LOS and NLOS scenarios, designing the illumination pattern and the RIS coefficients to minimize the reconstruction error of the reflectivity map in a desired area of interest. 

While the aforementioned works address the usage of RISs, whose phase pattern can be dynamically changed in time according to special needs, relevant research has been produced on non-reconfigurable metasurfaces, also called non-reconfigurable EM skins (NR-EMSs), whose wave manipulation capabilities are defined in the design process \cite{9975205,9580737}. NR-EMSs offer limited flexibility compared to RISs, but they come with orders-of-magnitude lower manufacturing costs, allowing potential mass-production of large devices \cite{9975205,9580737}. NR-EMS with conformal design have been targeted in our previous works for vehicular communication systems \cite{Mizmizi2022_conformal}, while few others addressed the potential of SP-EMS for localization (see \cite{tagliaferri2023reconfigurable,Lotti2023_metaprism}). 

\textit{Contribution}: This paper delves in the usage of NR-EMSs for radar imaging in NLOS in a way that multiples modules generates an artificial sweeping useful for imaging.
The selected use case is the high-resolution environment mapping for autonomous or assisted driving, where a radar-equipped traveling vehicle exploits an NR-EMS purposely deployed on a building to illuminate a desired area in NLOS ("\textit{see around the corner}"), and gathers the echoes undergoing double bounce with the NR-EMS. The peculiarity of such a system is that the vehicle, while moving, progressively illuminates different portions of the metasurface, namely \textit{modules}, each suitably designed to focus the impinging radiation on a desired spot in NLOS. The vehicle generates a synthetic aperture that maps into a \textit{multi-view} imaging system through the NR-EMS. The imaging resolution is dictated by the equivalent aperture on the NR-EMS, rather than the radar physical aperture. As the former can be very large, the proposed system allows high-resolution \textit{near-field} imaging with progressive low-resolution far-field acquisitions. 
We discuss the peculiar system design, outlining the related requirements. Simulation results confirm the feasibility of our imaging system and shed light on a promising usage of NR-EMSs for sensing.

\textit{Organization}: The paper is organized as follows: Sect. \ref{sect:multiview} describes the role of NR-EMSs in multi-view radar imaging, Sect. \ref{sect:system_model} outlines the system model, Sect. \ref{sect:system_design} details the design of the system and related trade-offs while Sect. \ref{sec:ImProcess} reports the image synthesis technique and the simulation results. Finally, Sect. \ref{sec:Conclusion} concludes the paper.

\textit{Notation}: The following notation is adopted in the paper: bold upper- and lower-case letters describe matrices and column vectors. Vector norm is denoted with $\|\mathbf{a}\|$. Vector transposition is $\mathbf{a}^T$. With $a\sim\mathcal{CN}(\mu,\sigma^2)$ we denote a multi-variate circularly complex Gaussian random variable $a$ with mean $\mu$ and variance $\sigma^2$. $\mathbb{E}[\cdot]$ is the expectation operator, while $\mathbb{R}$, $\mathbb{C}$ stand for the set of real and complex numbers, respectively. $\delta_{n}$ is the Kronecker delta function.

\section{NR-EMSs in Multi-view Radar Systems}\label{sect:multiview}
\begin{figure}[!b]
    \centering
    \subfloat[][]{\includegraphics[width=0.95\columnwidth]{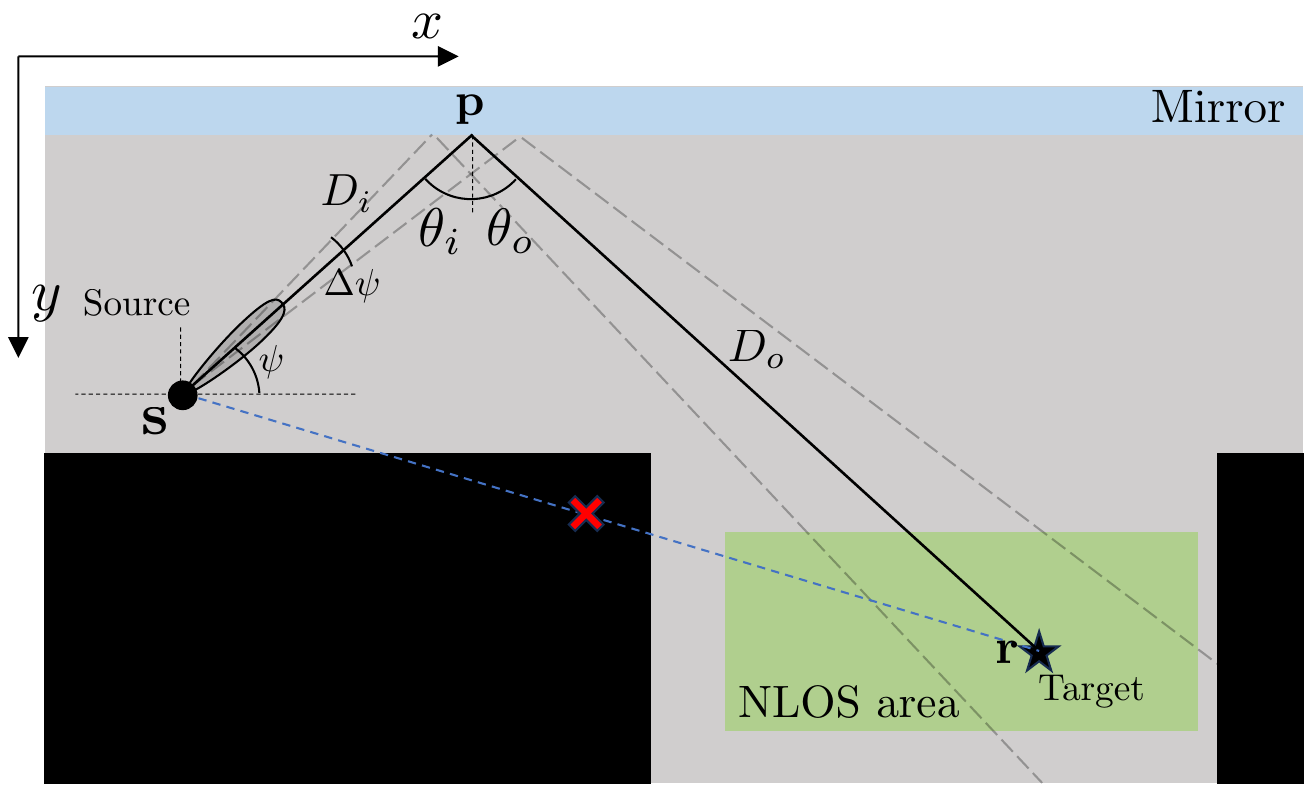}\label{subfig:mirror_1}}\\
    \subfloat[][]{\includegraphics[width=0.95\columnwidth]{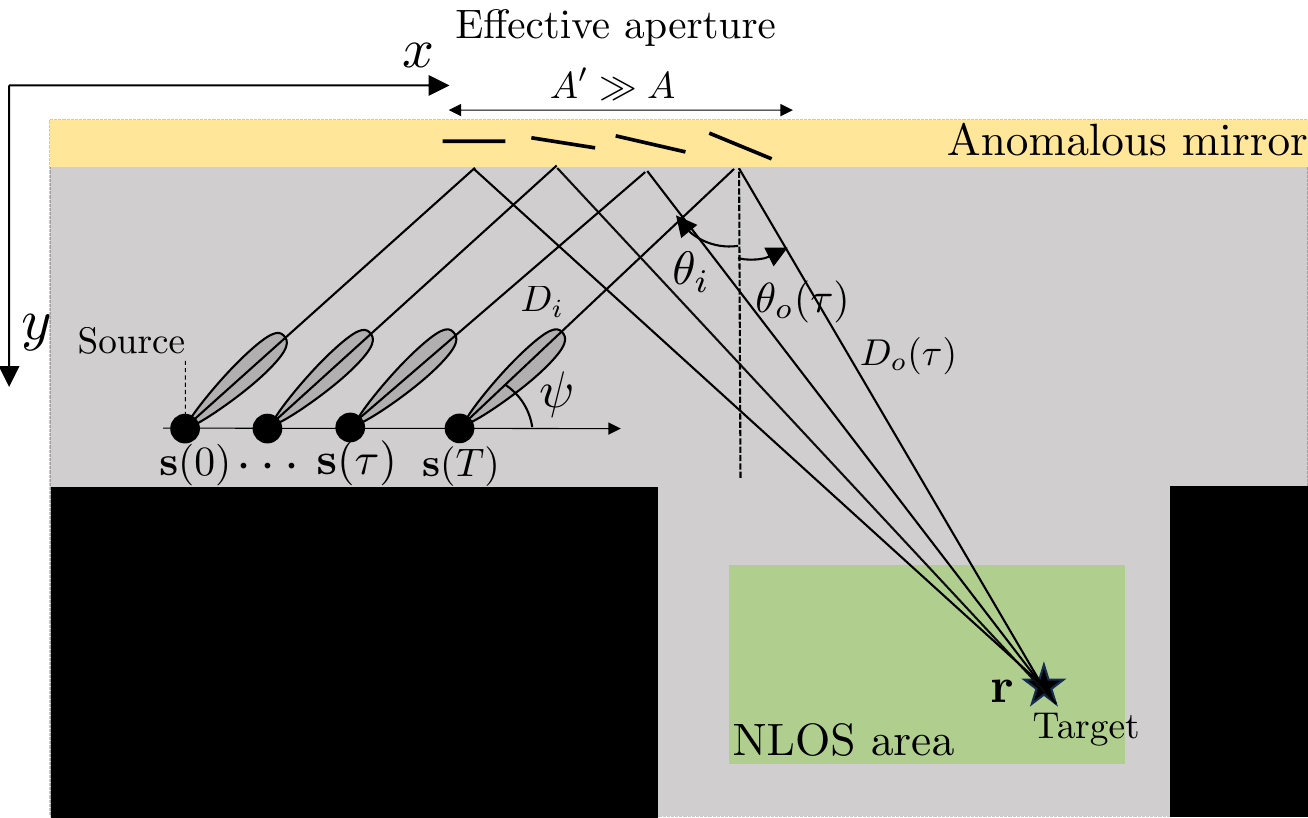}\label{subfig:EMS}}
    \caption{NLOS imaging with (a) reflection plane (mirror) and fixed source position, (b) NR-EMS and moving source.}
    \label{fig:multiview}
\end{figure}

Let us consider the 2D scenario depicted in Fig. \ref{subfig:mirror_1}, where a source equipped with a radar, located in $\mathbf{s}= [s_x, s_y]^T$, aims at creating an image of an area of interest where a target is located in $\mathbf{r} = [r_x, \,r_y]^T$. We assume that the area where the target is present is in NLOS, thus the image of the environment is enabled by the presence of a reflection plane deployed along $x$. The source is equipped with a radar with aperture $A$, emitting a signal along direction $\psi$ (pointing towards the reflection plane) over a narrow beamwidth $\Delta \psi \propto A^{-1}$. If the reflection plane is a mirror, the target is illuminated under specular reflection, i.e., for $\theta_i  = \theta_o$, where $\theta_i= \pi-\psi$ and $\theta_o$ are the incidence and reflection angles onto the reflection plane. The point $\mathbf{p}$ of specular reflection that allows imaging the target satisfies $s_y/D_i = r_y/D_o$ where $D_i=\| \mathbf{p}-\mathbf{s} \|$, $D_o=\| \mathbf{r}-\mathbf{p} \|$ denote the source-plane and plane-target distances, respectively. The resolution of the corresponding image of the target $\mathbf{r}$ is ruled by the radar's physical aperture $A$, thus reducing the beamwidth $\Delta \psi \propto A^{-1}$ yields better resolution.  

Since increasing the radar's physical aperture $A$ leads to a non-negligible additional cost for the hardware platform, we opt for a different solution to enhance the imaging resolution. Let us consider the case in which the radar moves along $x$ with velocity $v$ (see Fig. \ref{subfig:EMS}). The radar's position in space is now time-varying, $\mathbf{s}(\tau)$, where $\tau$ denotes the slow time. We assume to fix the pointing angle $\psi$ onto the reflection plane, thus the incident angle $\theta_i$ is fixed accordingly and the corresponding point on the plane is denoted by $\mathbf{p}(\tau) = [s_x(\tau) + D_i\cos\psi, 0]^T$. We have a \textit{multi-view} radar system if the reflection plane behaves like an anomalous mirror for the impinging radiation, namely it changes the reflection angle such that the signal emitted by the radar along its motion $\mathbf{s}(\tau)$, $\tau=0,...,T$ is \textit{focused} on the target in $\mathbf{r}$. The reflection angle at time $\tau$ shall be set as follows:
\begin{equation}
    \theta_o(\tau) = \arccos\left( \frac{r_y}{\| \mathbf{r} - \mathbf{p}(\tau)\|}\right)
\end{equation}
and the dependence on time can be related to a specific position onto the anomalous mirror through $\mathbf{p}(\tau)$. The system is depicted in Fig. \ref{subfig:EMS}. Now, with the aid of the anomalous mirror, the moving radar progressively increases the spatial diversity in the observation of the target, implement a \textit{synthetic aperture} that enhances both the resolution of the image of $\mathbf{r}$ as well as the SNR. Indeed, the image resolution is now ruled by the effective aperture onto the anomalous mirror: 
\begin{equation}\label{eq:aperture}
    A' \simeq vT + \frac{D_i \Delta \psi}{\sin\psi} \gg A,
\end{equation}
typically much larger that the aperture provided by the radar~\cite{9829750}. To cater to arbitrary geometries, we opt for the usage of a NR-EMS as the required anomalous mirror deployed on the reflection plane, whose design is detailed in the following.

\section{System Model}\label{sect:system_model}

The system model stems from Section \ref{sect:multiview}. The radar-equipped source moves along $x$, taking positions located in $\mathbf{s}(\tau) = [s_x(0)+v\tau, s_y]^T$, along the slow time interval $\tau\in[0,T]$. The radar periodically emits the pass-band signal 
\begin{equation}\label{eq:TX_signal}
    h(t) = g(t) e^{j2\pi f_0 t}
\end{equation}
where $t$ is the fast time, $g(t)$ is the base-band range-compressed pulse of bandwidth $B$, centered around carrier frequency $f_0$. Tx signal is emitted every pulse repetition interval $\Delta \tau$, that is herein neglected for the seek of simplicity in the notation. The radar is characterized by a linear aperture $A$ along $y$ (i.e., orthogonal to the direction of motion), that allows beamforming the pass-band signal \eqref{eq:TX_signal} towards direction $\psi$. The beamwidth on the azimuth plane (i.e., $xy$ plane) is $\Delta \psi \simeq \lambda_0/(A \cos \psi)$, where $\lambda_0=c/f_0$ is the carrier wavelength. 
The anomalous reflection is by a planar NR-EMS, made by $N$ $\lambda_0/4$-spaced elements along the $x$ axis , whose center is located in the origin of the global coordinate system\footnote{The 2D system model outlined in this paper is instrumental to describe the NR-EMS-aided radar imaging system design on the azimuth plane. In practice, planar NR-EMSs deployed on the $xz$ plane are required to guarantee a sufficient SNR. The 3D modeling would complicate the exposition and it is therefore not treated herein, but its development follows from the present with due adaptations. }.

At time instant $\tau$ the radar illuminates a portion of the NR-EMS made by
\begin{equation}
    N_{rad} \approx \frac{D_i \, \Delta \psi}{d \sin \psi} < N
\end{equation}
elements, for $n=n_0(\tau)-N_{rad}/2,...,n_0(\tau)+N_{rad}/2-1$, where $n_0(\tau)$ is the element index corresponding at the maximum pointing at time $\tau$, whose location in global coordinates is $\mathbf{x}_0(\tau)$.  
The general model of the received signal at the radar in $\mathbf{s}(\tau)$ due to scattering from target $\mathbf{r}$ through a double reflection from the NR-EMS is reported in \eqref{eq:Rx_sig_gen}.
\begin{figure*}[tb]
\begin{equation}
\label{eq:Rx_sig_gen}
\begin{split}
        y(t,\tau) = \sum_{n=n_0(\tau) - \frac{N_{rad}}{2}}^{n_0(\tau) + \frac{N_{rad}}{2}-1} 
        \sum_{n'=n_0(\tau)-\frac{N_{rad}}{2}}^{n_0(\tau)+\frac{N_{rad}}{2}-1} 
        \rho_{nn'}(\tau)
         \, e^{j \phi_n}e^{j\phi_{n'}} \,  g(t-\Delta_{nn'}(\tau)) \, e^{-j 2 \pi f_0 \Delta_{nn'}(\tau)}+ w(t,\tau)
\end{split}
\end{equation}
\hrulefill
\begin{equation}\label{eq:Rx_signal_generic_narrowbeam}
\begin{split}
        y(t,\tau) \overset{(a,b)}{\simeq} \rho (\tau) \,  g\left(t- \frac{2\left[D_i+D_o(\tau)\right]}{c}\right)& e^{-j \frac{4 \pi}{\lambda_0} \left[D_i + D_o(\tau)\right]} \times \\
        & \sum_{n=n_0(\tau) - \frac{N_{rad}}{2}}^{n_0(\tau) + \frac{N_{rad}}{2}-1} 
        \sum_{n'=n_0(\tau)-\frac{N_{rad}}{2}}^{n_0(\tau)+\frac{N_{rad}}{2}-1} 
        e^{j \phi_n}e^{j\phi_{n'}}  e^{-j \frac{2\pi d}{\lambda_0}(n+n') \left[ \sin\theta_i -\sin\theta_o(\tau)\right]}
         + w(t,\tau)
\end{split}
\end{equation}
\hrulefill
\end{figure*}
The two-way propagation delay between the source $\mathbf{s}(\tau)$ and target $\mathbf{r}$ is
\begin{equation}\label{eq:delay_NF}
    \Delta_{nn'}(\tau) = \frac{D_{\mathbf{s}n}(\tau)+D_{n\mathbf{r}}+D_{\mathbf{r}n'}+D_{n'\mathbf{s}}(\tau)}{c}
\end{equation}
accounting for propagation as forward scattering from the $n$th element and backward scattering from the $n'$th, $n'\neq n$ in general, $D_{\mathbf{s}n}(\tau) = \|\mathbf{x}_n - \mathbf{s}(\tau)\|$ being the distance between $\mathbf{s}(\tau)$ and the $n$th element in $\mathbf{x}_n$ (and similar definition for other terms), while $\phi_n$ is the NR-EMS phase at the $n$th element. The back-scattering power, whose expression is not reported for brevity, is inversely proportional to the product of the four travelled distances and proportional to the directivity of the radar, thus $|\rho_{nn'}(\tau)|^2 \propto D^{-2}_{\mathbf{s}n}(\tau) D^{-2}_{n\mathbf{r}} D^{-2}_{\mathbf{r}n'} D^{-2}_{n'\mathbf{s}}(\tau) \Delta \psi^{-4}$. Term $w(t,\tau)\in \mathcal{CN}(0, \sigma_w^2)$ is the additive white Gaussian noise corrupting the signal. Model \eqref{eq:Rx_sig_gen} describes a generic \textit{near-field} and \textit{spatially wideband} system, where the wavefront across the illuminated NR-EMS portion is curved and $\Delta_{nn'}(\tau) > 1/B$ in general.

\subsection{Simplified Narrow-Beam Model}\label{subsect:narrow_beam}
The usage of a comparatively narrow beam $\Delta \psi$ at the radar allows approximating the model of the Rx signal, yielding some useful simplifications. We have: $(a)$ \textit{far-field} when the wavefront across the portion of illuminated metasurface is approximated as planar and $(b)$ \textit{spatially narrowband} operation when the residual delays across the portion of illuminated metasurface do not practically affect the envelope of the Rx base-band pulse. 
Far-field implies 
\begin{equation}
\begin{split}\label{eq:delay_FF}
  \Delta_{nn'}(\tau) \simeq &\frac{2 D_i}{c} \hspace{-0.1cm}+ \hspace{-0.1cm}\frac{2 D_o(\tau)}{c}\hspace{-0.1cm} + \hspace{-0.1cm}(n\hspace{-0.1cm}+\hspace{-0.1cm}n') \frac{d}{c} [\sin \theta_i -\sin \theta_o(\tau)]
\end{split}
\end{equation}
where $D_i = \|\mathbf{x}_{0}(\tau) - \mathbf{s}(\tau)\| $ and $D_o(\tau) = \|\mathbf{r} - \mathbf{x}_{0}(\tau)\|$, 
and 
$|\rho(\tau)|^2\propto D^{-4}_i D^{-4}_o(\tau) \Delta\psi^{-4}$.
Spatially narrowband operation holds when
\begin{equation}
    \frac{1}{B} \gg N_{rad} \frac{d}{c} \cos \psi = \frac{D_i}{c} \Delta \psi \cot \psi,\,\,\,\, \psi \in (0,\pi/2]
\end{equation}
and it yields to this delayed echo:
\begin{equation}
    g(t-\Delta_{nn'}(\tau)) \simeq g\left(t- \frac{2 \left[D_i+D_o(\tau)\right]}{c} \right).
\end{equation}
The signal model under approximations $(a)$ and $(b)$ is reported in \eqref{eq:Rx_signal_generic_narrowbeam}.
While the far-field condition usually applies for beams $\Delta \psi$ in the order of some degrees and distances of few to tens of meters (as in the considered use case), the spatially narrowband operation is strongly dependent on the employed bandwidth. For instance, for $\psi=30$ deg, $D_i=10$ m, we obtain that $B < 100$ MHz for $\Delta \psi = 10$ deg, and $B<1$ GHz for $\Delta \psi = 1$ deg, the latter being the current limit for off-the-shelf automotive radars in terms of angular resolution~\cite{TI_ref_MMWCAS}.

\section{System Design}\label{sect:system_design}

The design of the system has the ultimate goal of enabling high-resolution imaging in NLOS conditions. If the radar employs a very large beamwidth, illuminating the whole NR-EMS, the NR-EMS phase profile is optimal only for a single position of the source and the target, as double focusing is required:
\begin{equation}
    \phi_n = \frac{2 \pi}{\lambda_0} \left(\big\|\mathbf{x}_n - \mathbf{s}\big\| + \big\|  \mathbf{r} - \mathbf{x}_n\big\|\right).
\end{equation}
Thus, the coverage is limited to a very small region in space, unless RISs are employed. In the following, we detail the system design using NR-EMSs under the narrow-beam approximation in Sect. \ref{subsect:narrow_beam}. 

\subsection{Modular NR-EMS}

The NR-EMS shall fulfill two goals, namely \textit{(i)} increasing the effective aperture of the radar, thus the resolution of the image, and \textit{(ii)} ensuring the full coverage of the NLOS area, i.e., any target shall be detectable within NLOS region. In this setting, the NR-EMS can be designed by composition of multiple equally sized \textit{modules}, each made by $N'\ll N$ elements (along $x$) and pre-configured with a linear phase profile on each. The latter eases the design and manufacturing of each module, and the optimal non-linear phase profile across the NR-EMS (for near-field focusing) is obtained by suitable configuration of multiple adjacent modules. Let us consider $L$ NR-EMS modules of $N'$ elements each. In order to enable the coherent combination of echoes from a target located in $\mathbf{r}$, (\textit{i}) the source must implement a synthetic aperture of duration $T \geq \frac{(LN' - N_{rad})d}{v}$ (assuming a narrow beamwidth, i.e., $N_{rad}  << L N' $) and (\textit{ii}) the $L$ modules must focus the impinging radiation from the location $\mathbf{s}(\tau)$ of the moving source to the location of the target, yielding a phase profile
\begin{equation}\label{eq:module_config}
    \phi_{n,\ell} = \frac{2 \pi d}{\lambda_0} n \left[\sin \left( \pi \hspace{-0.1cm} - \hspace{-0.1cm} \psi\right) -\sin \hspace{-0.1cm}\left( \arccos\hspace{-0.1cm}{\left(\frac{r_y}{\| \mathbf{r}-\mathbf{x}_{0,\ell}\|}\right)}\right)\right] 
\end{equation}
for $n=n_{0,\ell}-N'/2,...,n_{0,\ell}+N'/2-1$, $\ell=-L/2,...,L/2-1$, where $\mathbf{x}_{0,\ell}$ is the phase center of the $\ell$th module---identified by element $n_{0,\ell}$---illuminated by the moving source at time $\tau = \ell (N'd/v)$. With this configuration, the effective aperture of the radar system is \eqref{eq:aperture}, approximately equal to $A' \simeq L N' d$. Notice that this system allows near-field sensing by employing multiple NR-EMS modules configured for far-field, progressively approximating the curved wavefront with local planar ones.

\begin{figure}[!b]
    \centering
    \includegraphics[width=\columnwidth]{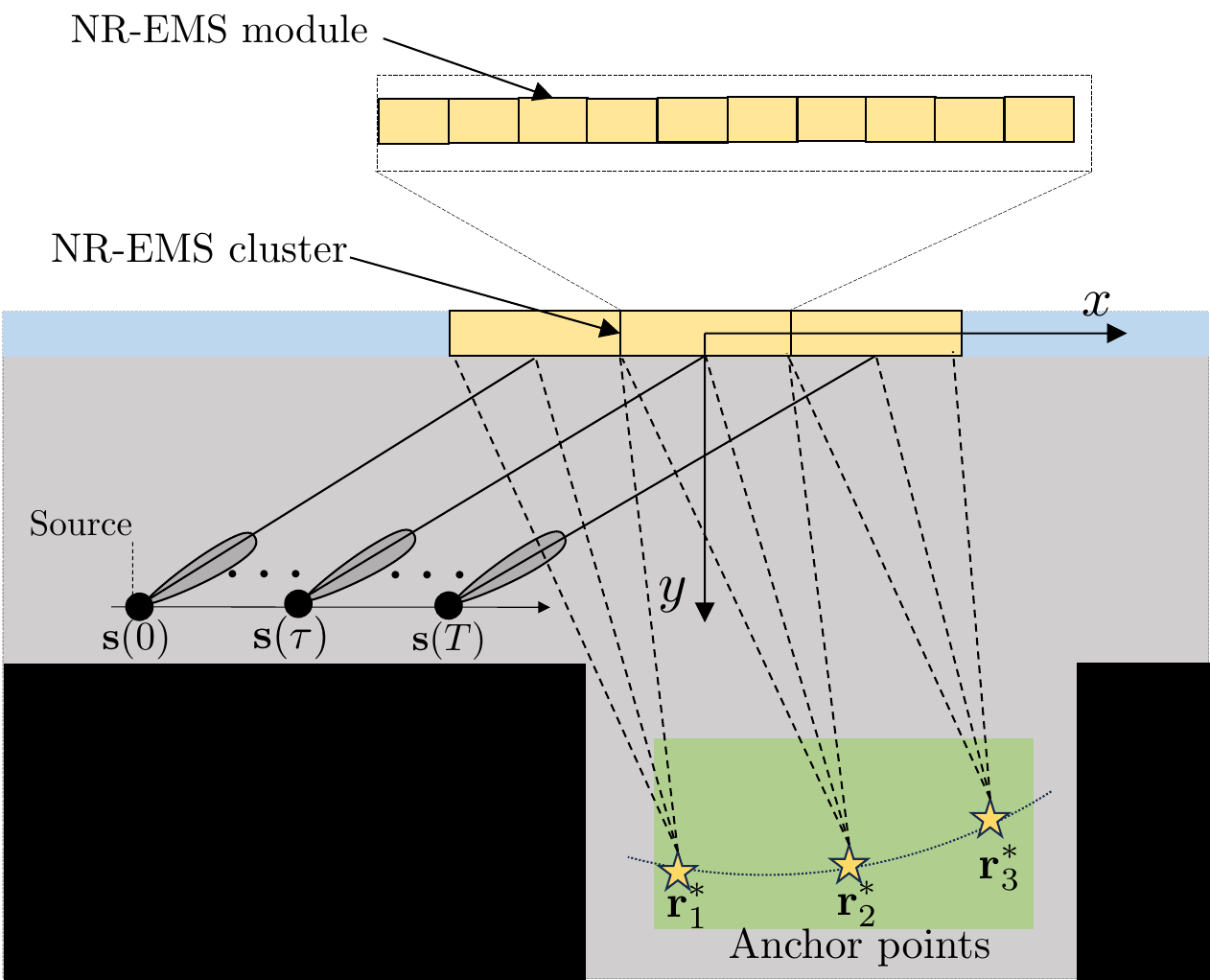}
    \caption{Modular NR-EMS for imaging in NLOS. The NR-EMS is made of multiple modules, organized in clusters. Each cluster focus the impinging radiation on a fixed anchor point, and each module implements a linear phase gradient.  }
    \label{fig:modular NR-EMS}
\end{figure}

With $L$ modules configured as \eqref{eq:module_config}, the impinging radiation from the radar is focused on a single point $\mathbf{r}$ in space, thus the system has enhanced resolution but not enough coverage. For the coverage of the entire NLOS area, it is necessary to consider multiple \textit{clusters} of $L$ modules, properly cascaded along $x$, each configured to focus on a different anchor point $\mathbf{r}^*$. By having $K$ clusters, a reasonable choice to allow the high-resolution imaging of the environment is to select the anchor points $\{\mathbf{r}_k^*\}_{k=1}^K$ over the iso-range curve at distance $R$ from the center of the NR-EMS. The result is depicted in Fig. \ref{fig:modular NR-EMS}. The phase gradient at the $\ell$th module of the $k$th cluster is obtained by \eqref{eq:module_config} plugging $\mathbf{r}=\mathbf{r}^*_k$. The complete image of the NLOS area is synthesized by illuminating \textit{all} the $K$ clusters through a synthetic aperture of duration $T \simeq KLN'/v$.


\subsection{Selection of the radar beamwidth}
The choice of the radar beamwidth $\Delta \psi$ is a degree of freedom for the proposed system. Employing a wide radar beam, such that $N_{\mathrm{rad}} \gg N'$, allows illuminating multiple modules at the same time instant $\tau$. Once the NR-EMS is deployed, a wide radar beamwidth relaxes the hardware requirements (and the cost) of the device, but it yields a lower SNR at the target's location compared to a narrow beamwidth. Possibly, if $N_{\mathrm{rad}} \simeq N$ (all the metasurface is illuminated in one shot), there is no need of a synthetic aperture. However, the SNR at the target's location diminishes by widening the radar beamwidth  $\Delta \psi$, as can be devised by inspection of the CRB on bistatic range estimation (see~\cite{Chetty2022_CRB} for further details and the analytical derivations)\footnote{The bistatic range considered in the CRB is the distance between the source (at the center of its synthetic aperture) to the phase center of the metasurface, plus the distance with the target.}. It can be shown that the relative SNR loss due to beam widening amounts to $\left(\frac{\Delta \psi_{\mathrm{wide}}}{\Delta \psi_{\mathrm{narrow}}}\right)^3$, pushing for the usage of the phased-array radars with the narrowest possible beamwidth compatible with the hardware availability. 

Besides the radar beamwidth, for $N_{\mathrm{rad}} < N$ it is necessary that the radar implements a further synthetic aperture to illuminate all the metasurface (all the $K$ clusters), to form the image without spatial ambiguities (\textit{ghost targets}). Those arise from partial illumination of multiple clusters that focus the radiation over different anchor points. If no synthetic aperture is implemented (thus no coherent combination of echoes from \textit{all} clusters), it is possible that ghost targets appear in the final image. Differently, by progressively illuminating the whole metasurface, ghost targets disappear.

\section{Image Synthesis and Simulation Results}
\label{sec:ImProcess}

\subsection{Image Synthesis}

The image is synthesized by back-projection (BP) in time, under narrow beamwidth assumption \eqref{eq:Rx_signal_generic_narrowbeam}. The BP technique for image synthesis is based on the wavefield migration integral extensively used in remote sensing to find the location of one or more targets~\cite{rs14153602}. It describes the spatio-temporal matched filtering (correlation) between the Rx signal and the model signal that would be received by back-scattering from a target in a given position in space (typically chosen over a pre-defined grid). The complex image value for a point $\mathbf{r}$ in the NLOS area can be computed as follows~\cite{rs14153602}\footnote{Image \eqref{eq:image_BP} can be also cast as an operation that provides the likelihood of the targets' positions in space, from which it follows their estimation by maximum likelihood approaches as described in \cite{5739256}.}:
\begin{equation}\label{eq:image_BP}
\begin{split}
    I(\mathbf{r}) \overset{(a,b)}{\simeq} \sum_{\tau} y\left(t = \frac{2 [D_i + D_o(\mathbf{r},\tau)]}{c}, \tau\right) e^{j \frac{4 \pi}{\lambda_0} [D_i + D_o(\mathbf{r},\tau)]}
\end{split}
\end{equation}
where $D_i= \| \mathbf{x}_{0}(\tau) - \mathbf{s}(\tau)\| = s_y/\sin\psi$ is the forward path toward the metasurface (that is constant over $\tau$)  while $D_o(\mathbf{r},\tau) = \| \mathbf{r} - \mathbf{x}_{0}(\tau)\|$ is the distance between the illuminated point at time $\tau$, i.e., $\mathbf{x}_{0}(\tau)$ and point $\mathbf{r}$. BP for narrow beamwidths and spatially wideband operation (over single slow-time snapshots) requires the relative position w.r.t. metasurface ($s_y$) as well as the pointing angle $\psi$.

\subsection{Simulation results}\label{sec:Res}

\begin{figure}[!t]
    \centering
    \includegraphics[width=\columnwidth]{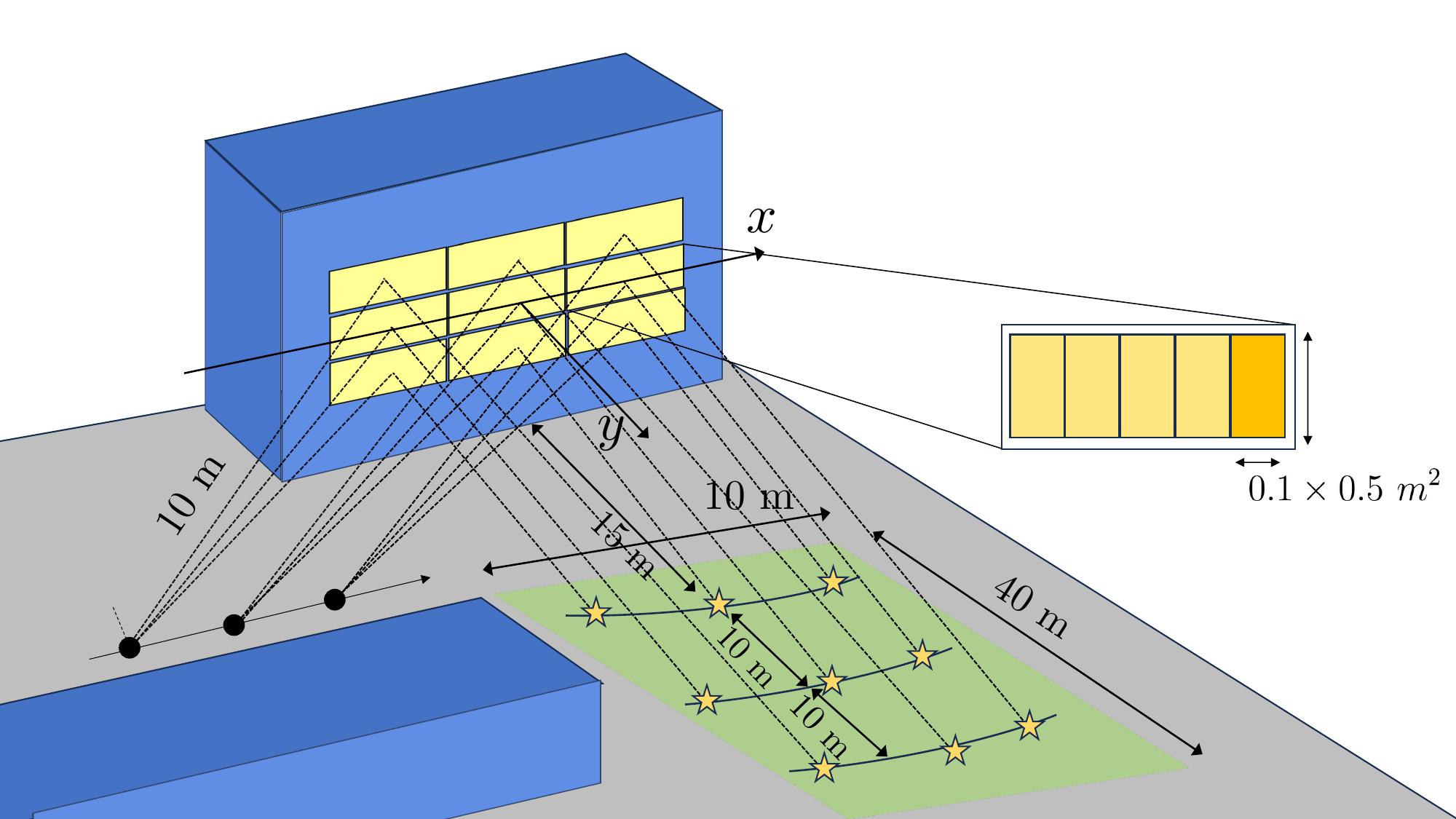}
    \caption{Sketch of the simulation environment with moving radar.}
    \label{fig:simulation_env}
\end{figure}

\begin{figure}[!t]
    \centering
    \includegraphics[width=.95\columnwidth]{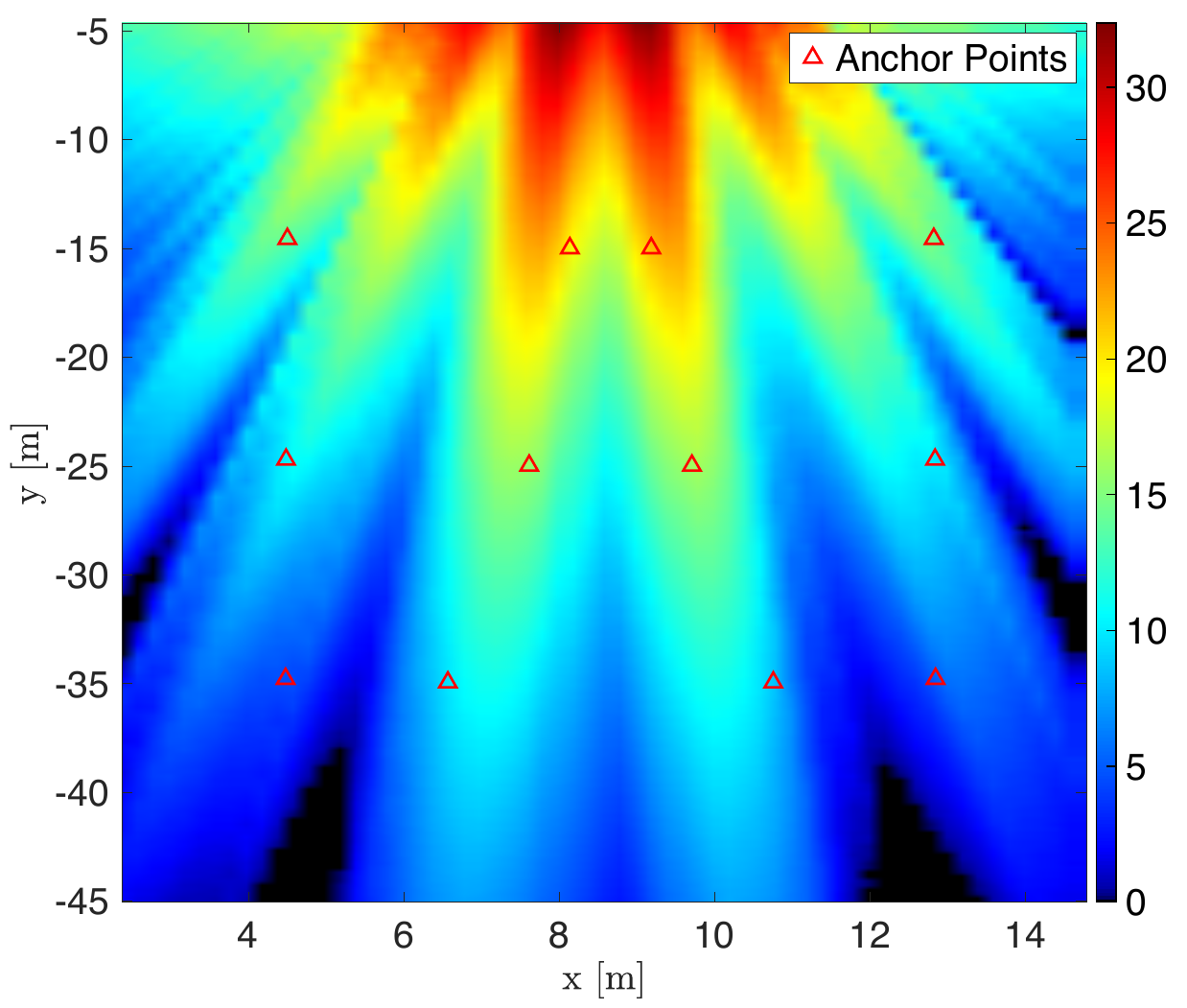}
    \caption{2D map of the SNR (in dB) in the NLOS area (image not in scale). The SNR is $>0$ dB in 98\% of the area.}
    \label{fig:SNR_map}
\end{figure}

\begin{figure}[!t]
    \centering
    \includegraphics[width=0.95\columnwidth]{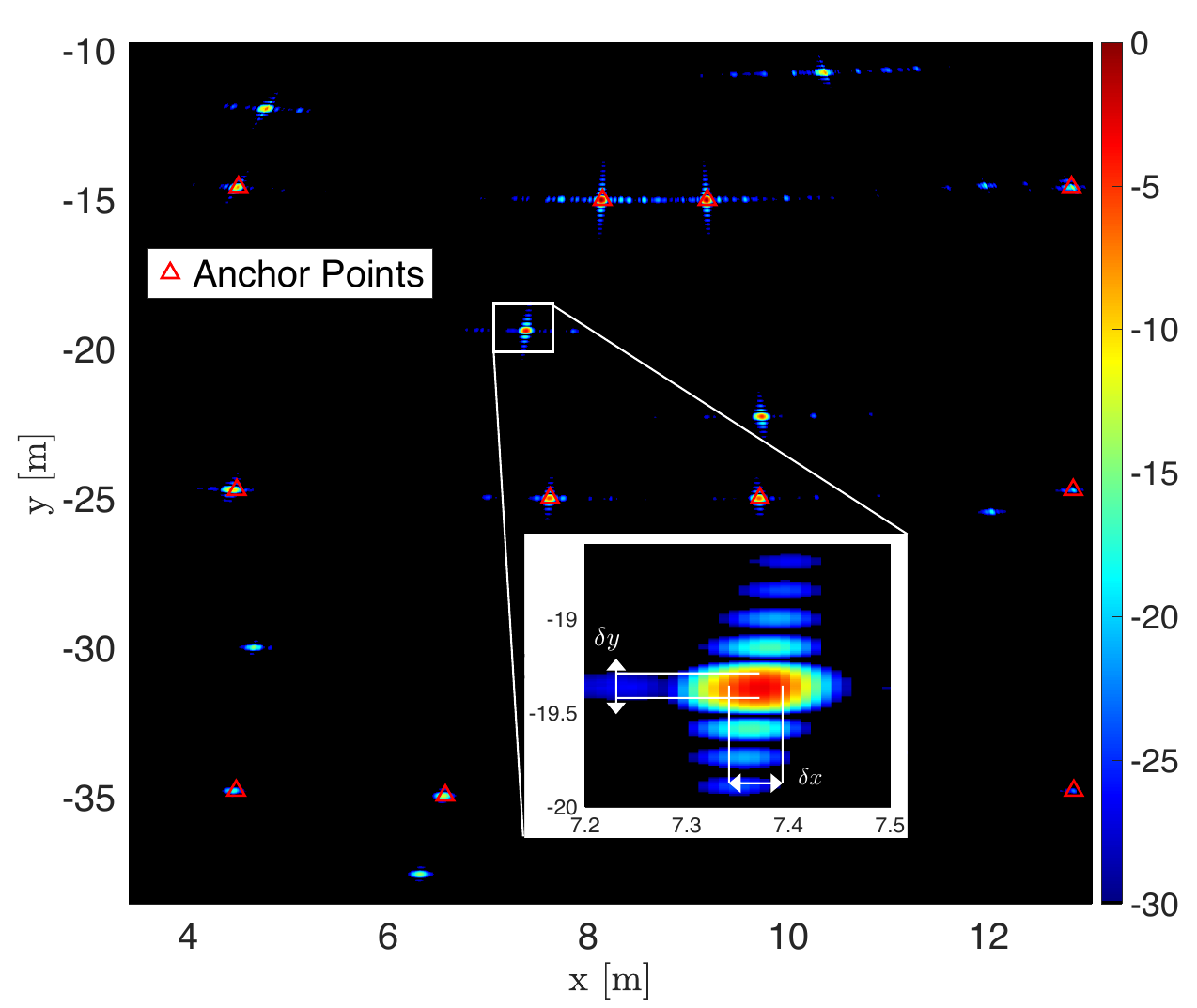}
    \caption{Focused image in the considered NLOS area, showing few point targets. The inset is a zoom on a selected target (images not in scale).}
    \label{fig:image}
\end{figure}

The system setup is skecthed in Fig. \ref{fig:simulation_env}. The simulation parameters are: carrier frequency $f_0=77$ GHz, employed bandwidth $B=1$ GHz, radar  beamwidth along the azimuth plane ($xy$) $\Delta \psi=1$ deg, Tx power $-10$ dBm. The NLOS area to be monitored is $10\times 40$ m$^2$. The NR-EMS is made by $K = 4$ clusters of $L = 5$ modules, each of size $0.1$ m along $x$ (horizontal) and $0.5$ m along $z$ (vertical). We assume that the radar's physical aperture is 1D, as for a uniform linear array, thus the elevation beamwidth is large\footnote{As for typical automotive radars, achieving high azimuth resolution and no (or poor) elevation resolution \cite{TI_ref_MMWCAS}.}. In this latter setting, we exploit the vertical direction on the reflection plane to deploy additional NR-EMS clusters (up to 12) to increase the number of anchor points on the NLOS covered area. The results are discussed in the following.

Fig. \ref{fig:SNR_map} reports the achieved SNR map (in dB scale) in the NLOS area. The 12 anchor points are selected to be deployed at range distances of $15,25,35$ m from the phase center of the NR-EMS. The synthetic aperture is $2$ m such that to illuminate all the NR-EMS clusters. The SNR map is built by deploying regularly spaced fictitious targets with an RCS of $0.1$ m$^2$ and evaluate the SNR after image formation. In the considered settings, the proposed system is able to achieve an SNR above 0 dB in $\simeq 98\%$ of the NLOS area, allowing the detection of targets notwithstanding the severe path-loss induced by the double reflection on the NR-EMS. It is worth remarking that such an SNR can be attained with very large metasurfaces, that can be practically manufactured only as NR-EMS. 

The second remarkable result, reported in Fig. \ref{fig:image}, shows instead a focused radar image of a few randomly deployed point targets in the NLOS area, each with an RCS of $0.5$ m$^2$. The image is normalized to the maximum value and reported in dB scale. Remarkably, we have that \textit{(i)} both targets in near and far range are detectable (as the SNR after image formation is comparably large) and \textit{(ii)} each point target is imaged with a resolution of few to tens of centimeters, with little variation in space. The latter result should be compared with the radar's resolution capabilities without the NR-EMS focusing effect. For instance, for the target in the inset of Fig. \ref{fig:image}, located at approx $D_o = 20$ m from the NR-EMS and $D_i= 10$ m from the source, the native cross-range resolution of the radar (approximately equivalent to the resolution along $x$) would be $\delta x \simeq (D_i+D_o)\Delta \psi = 50$ cm, while drops to $\delta x \simeq 10$ cm thanks to the NR-EMS. The range resolution (circa $\delta y$) is instead ruled by the employed bandwidth for the considered point $\delta y\simeq c/2B = 15$ cm, with a slight improvement when using the NR-EMS thanks to the near-field effect. The results show the feasibility of the proposed imaging system with NR-EMS, and shed light on a promising usage of passive metasurfaces for NLOS sensing.

\section{Conclusions}
\label{sec:Conclusion}

This work proposes and discusses a novel radar system exploiting a modular low-cost NR-EMS for NLOS imaging of a desired area in the environment, with application to automotive and road safety. The proposed system leverages the vehicle's motion to synthesize an aperture on the NR-EMS that allows for increasing the physical radar's aperture, enabling high-resolution imaging in NLOS over a desired area of interest. Design peculiarities are discussed and simulation results are presented to validate the idea. Next investigations will involve generalization to integrated communication and sensing over NR-EMS, and further simulations for more realistic settings with possibly dedicated experimentation.

\bibliographystyle{IEEEtran}
\bibliography{Bibliography}

\begin{thebibliography}{10}
\providecommand{\url}[1]{#1}
\csname url@samestyle\endcsname
\providecommand{\newblock}{\relax}
\providecommand{\bibinfo}[2]{#2}
\providecommand{\BIBentrySTDinterwordspacing}{\spaceskip=0pt\relax}
\providecommand{\BIBentryALTinterwordstretchfactor}{4}
\providecommand{\BIBentryALTinterwordspacing}{\spaceskip=\fontdimen2\font plus
\BIBentryALTinterwordstretchfactor\fontdimen3\font minus
  \fontdimen4\font\relax}
\providecommand{\BIBforeignlanguage}[2]{{%
\expandafter\ifx\csname l@#1\endcsname\relax
\typeout{** WARNING: IEEEtran.bst: No hyphenation pattern has been}%
\typeout{** loaded for the language `#1'. Using the pattern for}%
\typeout{** the default language instead.}%
\else
\language=\csname l@#1\endcsname
\fi
#2}}
\providecommand{\BIBdecl}{\relax}
\BIBdecl

\bibitem{doi:10.1126/science.1210713}
\BIBentryALTinterwordspacing
N.~Yu, P.~Genevet, M.~A. Kats, F.~Aieta, J.-P. Tetienne, F.~Capasso, and
  Z.~Gaburro, ``Light propagation with phase discontinuities: Generalized laws
  of reflection and refraction,'' \emph{Science}, vol. 334, no. 6054, pp.
  333--337, 2011. [Online]. Available:
  \url{https://www.science.org/doi/abs/10.1126/science.1210713}
\BIBentrySTDinterwordspacing

\bibitem{Buzzi_RISforradar_journal}
S.~Buzzi, E.~Grossi, M.~Lops, and L.~Venturino, ``Foundations of mimo radar
  detection aided by reconfigurable intelligent surfaces,'' \emph{IEEE
  Transactions on Signal Processing}, vol.~70, pp. 1749--1763, 2022.

\bibitem{9511765}
A.~Aubry, A.~De~Maio, and M.~Rosamilia, ``Ris-aided radar sensing in n-los
  environment,'' in \emph{2021 IEEE 8th International Workshop on Metrology for
  AeroSpace (MetroAeroSpace)}, 2021, pp. 277--282.

\bibitem{9508872}
A.~Elzanaty, A.~Guerra, F.~Guidi, and M.-S. Alouini, ``Reconfigurable
  intelligent surfaces for localization: Position and orientation error
  bounds,'' \emph{IEEE Transactions on Signal Processing}, vol.~69, pp.
  5386--5402, 2021.

\bibitem{9838638}
Z.~Wang, Z.~Liu, Y.~Shen, A.~Conti, and M.~Z. Win, ``Source localization with
  intelligent surfaces,'' in \emph{ICC 2022 - IEEE International Conference on
  Communications}, 2022, pp. 895--900.

\bibitem{9625826}
D.~Dardari, N.~Decarli, A.~Guerra, and F.~Guidi, ``Los/nlos near-field
  localization with a large reconfigurable intelligent surface,'' \emph{IEEE
  Transactions on Wireless Communications}, vol.~21, no.~6, pp. 4282--4294,
  2022.

\bibitem{9650561}
M.~Luan, B.~Wang, Y.~Zhao, Z.~Feng, and F.~Hu, ``Phase design and near-field
  target localization for ris-assisted regional localization system,''
  \emph{IEEE Transactions on Vehicular Technology}, vol.~71, no.~2, pp.
  1766--1777, 2022.

\bibitem{PhysRevLett.92.193904}
\BIBentryALTinterwordspacing
G.~Lerosey, J.~de~Rosny, A.~Tourin, A.~Derode, G.~Montaldo, and M.~Fink, ``Time
  reversal of electromagnetic waves,'' \emph{Phys. Rev. Lett.}, vol.~92, p.
  193904, May 2004. [Online]. Available:
  \url{https://link.aps.org/doi/10.1103/PhysRevLett.92.193904}
\BIBentrySTDinterwordspacing

\bibitem{Ghavami2022_metasurfaces_radar}
N.~Ghavami, E.~Razzicchia, O.~Karadima, P.~Lu, W.~Guo, I.~Sotiriou, E.~Kallos,
  G.~Palikaras, and P.~Kosmas, ``The use of metasurfaces to enhance microwave
  imaging: Experimental validation for tomographic and radar-based
  algorithms,'' \emph{IEEE Open Journal of Antennas and Propagation}, vol.~3,
  pp. 89--100, 2022.

\bibitem{9299878}
Y.~Tao and Z.~Zhang, ``Distributed computational imaging with reconfigurable
  intelligent surface,'' in \emph{2020 International Conference on Wireless
  Communications and Signal Processing (WCSP)}, 2020, pp. 448--454.

\bibitem{jiang2023near}
Y.~Jiang, F.~Gao, Y.~Liu, S.~Jin, and T.~Cui, ``Near field computational
  imaging with ris generated virtual masks,'' 2023.

\bibitem{torcolacci2023holographic}
G.~Torcolacci, A.~Guerra, H.~Zhang, F.~Guidi, Q.~Yang, Y.~C. Eldar, and
  D.~Dardari, ``Holographic imaging with xl-mimo and ris: Illumination and
  reflection design,'' 2023.

\bibitem{9975205}
G.~Oliveri, F.~Zardi, P.~Rocca, M.~Salucci, and A.~Massa, ``Constrained design
  of passive static em skins,'' \emph{IEEE Transactions on Antennas and
  Propagation}, vol.~71, no.~2, pp. 1528--1538, 2023.

\bibitem{9580737}
G.~Oliveri, P.~Rocca, M.~Salucci, and A.~Massa, ``Holographic smart em skins
  for advanced beam power shaping in next generation wireless environments,''
  \emph{IEEE Journal on Multiscale and Multiphysics Computational Techniques},
  vol.~6, pp. 171--182, 2021.

\bibitem{Mizmizi2022_conformal}
M.~Mizmizi, R.~A. Ayoubi, D.~Tagliaferri, K.~Dong, G.~G. Gentili, and
  U.~Spagnolini, ``Conformal metasurfaces: a novel solution for vehicular
  communications,'' \emph{IEEE Transactions on Wireless Communications}, pp.
  1--1, 2022.

\bibitem{tagliaferri2023reconfigurable}
D.~Tagliaferri, M.~Mizmizi, G.~Oliveri, U.~Spagnolini, and A.~Massa,
  ``Reconfigurable and static em skins on vehicles for localization,'' 2023.

\bibitem{Lotti2023_metaprism}
M.~Lotti and D.~Dardari, ``Metaprism-aided nlos target localization,'' in
  \emph{2023 31st European Signal Processing Conference (EUSIPCO)}, 2023, pp.
  895--899.

\bibitem{9829750}
G.~Lerosey and M.~Fink, ``Wavefront shaping for wireless communications in
  complex media: From time reversal to reconfigurable intelligent surfaces,''
  \emph{Proceedings of the IEEE}, vol. 110, no.~9, pp. 1210--1226, 2022.

\bibitem{TI_ref_MMWCAS}
T.~Instruments, ``Imaging radar using cascaded mmwave sensor reference
  design,'' available at: https://www.ti.com/lit/ug/tiduen5a/tiduen5a.pdf.

\bibitem{Chetty2022_CRB}
A.~Liu, Z.~Huang, M.~Li, Y.~Wan, W.~Li, T.~X. Han, C.~Liu, R.~Du, D.~K.~P. Tan,
  J.~Lu, Y.~Shen, F.~Colone, and K.~Chetty, ``A survey on fundamental limits of
  integrated sensing and communication,'' \emph{IEEE Communications Surveys \&
  Tutorials}, vol.~24, no.~2, pp. 994--1034, 2022.

\bibitem{rs14153602}
\BIBentryALTinterwordspacing
S.~Tebaldini, M.~Manzoni, D.~Tagliaferri, M.~Rizzi, A.~V. Monti-Guarnieri,
  C.~M. Prati, U.~Spagnolini, M.~Nicoli, I.~Russo, and C.~Mazzucco, ``Sensing
  the urban environment by automotive sar imaging: Potentials and challenges,''
  \emph{Remote Sensing}, vol.~14, no.~15, 2022. [Online]. Available:
  \url{https://www.mdpi.com/2072-4292/14/15/3602}
\BIBentrySTDinterwordspacing

\bibitem{5739256}
J.~Gunther, R.~West, N.~Crookston, and T.~Moon, ``Maximum likelihood synthetic
  aperture radar image formation for highly nonlinear flight tracks,'' in
  \emph{2011 Digital Signal Processing and Signal Processing Education Meeting
  (DSP/SPE)}, 2011, pp. 449--454.

\end{thebibliography}


\begin{thebibliography}{00}
\bibitem{b1} M. Di Renzo et al., ``Smart Radio Environments Empowered by Reconfigurable Intelligent Surfaces: How It Works, State of Research, and The Road Ahead,'' in IEEE Journal on Selected Areas in Communications, vol. 38, no. 11, pp. 2450-2525, Nov. 2020, doi: 10.1109/JSAC.2020.3007211.
\bibitem{b2} M. A. ElMossallamy et al., ``Reconfigurable Intelligent Surfaces for Wireless Communications: Principles, Challenges, and Opportunities,'' in IEEE Transactions on Cognitive Communications and Networking, vol. 6, no. 3, pp. 990-1002, Sept. 2020, doi: 10.1109/TCCN.2020.2992604.
\bibitem{b4} D. Tagliaferri, et al. "Reconfigurable and Static EM Skins on Vehicles for Localization." arXiv preprint arXiv:2308.04319 (2023).
\bibitem{b7} M. Mizmizi et al., "Conformal Metasurfaces for Recovering Dynamic Blockage in Vehicular Systems," GLOBECOM 2022 - 2022 IEEE Global Communications Conference, Rio de Janeiro, Brazil, 2022, pp. 6025-6030, doi: 10.1109/GLOBECOM48099.2022.10001090.
\bibitem{b8} S. Buzzi et al., "Foundations of MIMO Radar Detection Aided by Reconfigurable Intelligent Surfaces," in IEEE Transactions on Signal Processing, vol. 70, pp. 1749-1763, 2022, doi: 10.1109/TSP.2022.3157975.
\bibitem{b9} S. Buzzi et al., "RIS-Aided Monostatic Mimo Radar with Co-Located Antennas," ICASSP 2022 - 2022 IEEE International Conference on Acoustics, Speech and Signal Processing (ICASSP), Singapore, Singapore, 2022, pp. 4998-5002, doi: 10.1109/ICASSP43922.2022.9747807.
\bibitem{b10} A. Aubry, A. De Maio and M. Rosamilia, "Reconfigurable Intelligent Surfaces for N-LOS Radar Surveillance," in IEEE Transactions on Vehicular Technology, vol. 70, no. 10, pp. 10735-10749, Oct. 2021, doi: 10.1109/TVT.2021.3102315.
\bibitem{b18} Nanfang Yu et al. ,Light Propagation with Phase Discontinuities: Generalized Laws of Reflection and Refraction.Science334,333-337(2011).DOI:10.1126/science.1210713
\bibitem{b20} U. Spagnolini, ''Statistical Signal Processing in Engineering". John Wiley \& Sons Ltd, 2018.
\bibitem{b14} D. Tagliaferri et al., "Navigation-Aided Automotive SAR for High-Resolution Imaging of Driving Environments," in IEEE Access, vol. 9, pp. 35599-35615, 2021, doi: 10.1109/ACCESS.2021.3062084.
\end{thebibliography}

\end{document}